\documentclass[prd,aps,superscriptaddress,nofootinbib]{revtex4}
\usepackage{graphicx}
\usepackage{amsfonts}

\usepackage{amsfonts}
\usepackage{amssymb}
\usepackage{amsmath}
\usepackage{dsfont}
\usepackage{hyperref}
\usepackage{slashed}
\usepackage{empheq}
\usepackage{multirow}
\usepackage{fancyhdr}
\usepackage{appendix}
\usepackage{subfigure}
\usepackage{times}

\DeclareMathOperator{\hz}{h\relax{\kern-.15em}z}
\DeclareMathOperator{\pz}{\psi\relax{\kern-.15em}z}

\usepackage[english]{babel}
\usepackage{amssymb}
\usepackage{amsfonts}
\usepackage{amsmath}
\usepackage{graphicx}
\usepackage{epsfig}
\usepackage{psfrag}
\newcommand{\be}{\begin{equation}} \newcommand{\ee}{\end{equation}}
\newcommand{\bea}{\begin{eqnarray}} \newcommand{\eea}{\end{eqnarray}}
\newcommand{\beann}{\begin{eqnarray*}}  \newcommand{\eeann}{\end{eqnarray*}}
\newcommand{\bfig}{\begin{figure}} \newcommand{\efig}{\end{figure}}
\newcommand{\ba}{\begin{array}} \newcommand{\ea}{\end{array}}
\newcommand{\bcen}{\begin{center}} \newcommand{\ecen}{\end{center}}
\newcommand{\btab}{\begin{tabular}} \newcommand{\etab}{\end{tabular}}
\newcommand{\nn}{\nonumber}
\newcommand{\matt}{\left ( \begin{array}{ccc}}
    \newcommand{\ematt}{\end{array} \right )} \newcommand{\matf}{\left ( \begin{array}{cccc}}
    \newcommand{\ematf}{\end{array} \right )} \newcommand{\vect}{\left ( \begin{array}{c}}
    \newcommand{\evect}{\end{array} \right )}    \def\beqn{\begin{eqnarray}}
 \def\eeqn{\end{eqnarray}}  
     
\newtheorem{Proposition}{Proposition}[section]

\newtheorem{Theorem}{Theorem}[section]
\newtheorem{Lemma}{Lemma}[section]
\newtheorem{Corrolary}{Corrolary}[section]

\newcommand{\bp}{\begin{Proposition}}   \newcommand{\ep}{\end{Proposition}}
\newcommand{\bt}{\begin{Theorem}}   \newcommand{\et}{\end{Theorem}}
\newcommand{\bl}{\begin{Lemma}}     \newcommand{\el}{\end{Lemma}}
\newcommand{\bc}{\begin{Corrolary}} \newcommand{\ec}{\end{Corrolary}}

\begin{document}

\title{Charged Proca Stars}

\author{Ignacio Salazar Landea}\email{peznacho@gmail.com}
\affiliation{ Centro At\'omico Bariloche, 8400-S.C. de Bariloche,
R\'{\i}o Negro, Argentina}

\author{Federico Garc\'ia}\email{fgarcia@iar-conicet.gov.ar}
\affiliation{Instituto Argentino de Radioastronom\'{\i}a (CCT La Plata, CONICET), C.C.5, 1894 Villa Elisa, Argentina\\
Facultad de Ciencias Astron\'omicas y Geof\'{\i}sicas, Universidad Nacional de La Plata, Paseo del Bosque, B1900FWA La Plata, Argentina}

\date{\today}
\begin{abstract}
In this paper, we study gauged solutions associated to a massive vector field representing a spin-one condensate, namely the Proca field. We focus on regular spherically-symmetric solutions which we construct either using a self-interaction potential or general relativity in order to glue the solutions together. 
We start generating non-gravitating solutions, so-called, Proca Q-balls and charged Proca Q-balls. Then, we turn-on backreaction on the metric, allowing gravity to hold together the Proca condensate, to study the so-called Proca stars, charged Proca stars, Proca Q-stars and charged Proca Q-stars.
\end{abstract}

\pacs{ }

\maketitle

\section{Introduction}

Boson stars are gravitationally bounded macroscopic states made up of bosons. Firstly introduced in \cite{Kaup:1968zz, Ruffini:1969qy} as spherically symmetric solutions of the Einstein-Klein-Gordon equations, they remain being objects of study until present. Of particular interest are their possible astrophysical applications running from black-hole mimickers \cite{Meliani:2015zta,Grandclement:2014msa,Vincent:2015xta,Cunha:2016bpi}, black hole hair and clouds \cite{Hod:2012px,Hod:2014baa,Benone:2014ssa,Herdeiro:2014pka,Herdeiro:2014goa,Sampaio:2014swa,Herdeiro:2016tmi} to dark matter candidates \cite{DM}. 
Being simple tractable objects, they serve also as toy models helpful to understand the interplay between Field Theory, General Relativity and its possible extensions \cite{Torres:1997np}.

In order to construct boson stars, a fundamental aspect to be considered is the existence of an internal $U(1)$ symmetry, which will introduce an associated conserved charge that may give rise to classical solutions with non-zero total charge. Thus, a natural step forward is to gauge the global $U(1)$ symmetry. This was first done in \cite{Jetzer:1989av}, giving birth to the so-called charged boson stars.

An important family of classical objects was introduced in \cite{Coleman:1985ki}, where regular solutions for the non-gravitating Klein-Gordon equation of motion were found, known as Q-balls. These objects also rely on the existence of a conserved charge and can be obtained through the minimization of their corresponding free energy. Since gravity does not glue this solutions together, the parameters of their self-interaction potentials must have the right values in order to achieve actual solutions. Some interesting general properties were presented in \cite{Gulamov:2013cra,Gulamov:2015fya}. Gauged extensions to these particular class were presented in \cite{Lee:1988ag}. 

When coupled to gravity, Q-balls become Q-stars \cite{Lynn:1988rb,Prikas:2002ij}. Several other potentials, that would not give regular spherical solutions if not coupled to gravity, but of theoretical importance in other field theoretical contexts, were also studied under this framework \cite{Colpi:1986ye,Hartmann:2012da,Kumar:2016oop,Kumar:2015sia,Schunck:1999zu}.

Recently, new models of boson stars and Q-balls where presented in \cite{Loginov:2015rya,Brito:2015pxa}. In these works, a vector field representing a spin-one condensate is studied, namely the Proca field. Hence, those particular solutions are the Proca stars and Proca-balls, respectively. In this paper, we study gauged extensions to these models, so-called charged Proca-balls and stars. For completeness we also consider self-interacting gauged and non-gauged gravitating solutions, which we shall call Proca Q-stars. 

 Real vector solutions were also studied in the literature \cite{Brito:2015yga,Brito:2015yfh} in the context of dark matter cores and accretion in compact objects. Without an internal symmetry this solutions can not be static and correspond to the vector equivalent of the scalar oscillantons introduced in \cite{Seidel:1991zh}.

The paper organizes as follows: in Section~\ref{smodel} we present our general model, in Section~\ref{sballs} we study non-gravitating solutions, i.e. Proca Q-balls and charged Proca Q-balls and, in Section~\ref{stars}, we turn-on backreaction on the metric, allowing gravity to glue together the Proca condensate, giving rise to solutions named: Proca stars, charged Proca stars, Proca Q-stars and charged Proca Q-stars. Finally, in Section~\ref{conc} we summarize our results and elucidate some possible future directions.

\section{The model}
\label{smodel}

By means of introducting a vector field representing a spin-one condensate, namely the Proca field, we arrive to the Maxwell-Einstein-Proca model, whose action reads
\be
S=\int d^4x \sqrt{-g}\left( R-\frac12 \bar{B}^{\mu\nu} {B}_{\mu\nu} -U\left( \bar{B}^{\mu} {B}_{\mu} \right)- \frac14 F^{\mu\nu}F_{\mu\nu}     \right)\,,
\label{action}
\ee
where $F_{\mu\nu}=\nabla_\mu A_\nu - \nabla_\nu A_\mu  $ is the $U(1)$ gauge-field strength and  $B_{\mu\nu}=D_\mu B_\nu -D_\nu B_\mu$ with $D_\mu=\nabla_\mu-i q A_\mu$. A similar lagrangian with negative cosmological constant was recently studied in \cite{Cai:2013aca,Cai:2013pda} in the context
 of AdS/CFT duality as a model for holographic p-wave superfluids
\footnote{A non-minimal 
coupling term $i q \gamma B_\mu \bar{B}_\nu F^{\mu\nu}$  characterizing the magnetic moment of the vector field could be an interesting extension to this model (see, for instance, \cite{Cai:2013pda, Sampaio:2014swa}).}.

Varying the action introduced above (\ref{action}) we derive the following  equations of motion 
\be
D^{\nu}B_{\nu\mu} - U'\left( \bar{B}^{\mu} {B}_{\mu} \right) =0\,,
\label{proca}
\ee
\be
\nabla^{\nu}F_{\nu\mu}-i q \left( B^{\nu} \bar{B}_{\nu\mu} - \bar{B}^{\nu} {B}_{\nu\mu}  \right)=0\,,
\label{maxwell}
\ee
\be
R_{\mu\nu}-\frac12 R g_{\mu\nu}= T^{(B)}_{\mu\nu}+T^{(F)}_{\mu\nu}\,,
\label{einstein}
\ee
where $R_{\mu\nu}$ is the Ricci tensor and $T_{\mu\nu}$ the energy-momentum tensor of the matter fields.

The energy-momentum tensor has two different components. The first one, associated to the Proca field reads
\be
T^{(B)}_{\mu\nu}=-{B}_{\mu\lambda} \bar{B}_{\nu}^{\lambda} - \bar{B}_{\mu\lambda} {B}_{\nu}^{\lambda} +\frac12 g_{\mu\nu} B_{\sigma\lambda} \bar{B}^{\sigma\lambda}-U'\left( \bar{B}^{\mu} {B}_{\mu} \right) \left(B_\mu\bar{B}_\nu + \bar{B}_\mu{B}_\nu \right) + U\left( \bar{B}^{\mu} {B}_{\mu} \right) g_{\mu\nu} \,.
\ee
while, the second one, associated to the Maxwell field is
\be
T^{(F)}_{\mu\nu}=-{F}_{\mu\lambda}{F}_{\nu}^{\lambda} +\frac12 g_{\mu\nu} F_{\sigma\lambda} F^{\sigma\lambda}\,.
\ee

Since the model is invariant under $U(1)$, a Noether current associated to this symmetry arises. This current is defined by
\be
j^\mu=\frac12  \left(  \bar{B}^{\mu\nu} B_\nu -{B}^{\mu\nu} \bar{B}_\nu  \right)\,.
\ee

\section{Charged Proca balls}
\label{sballs}

In this section we study non-gravitationally bounded solutions commonly named as ``balls''  \cite{Coleman:1985ki} in the literature. For this purpose, we consider a fixed Minkowski metric associated to a background flat space given by

\be
ds^2=-dt^2+dr^2+r^2 d\Omega_2^2 \,.
\ee

In order to achieve actual solutions for these non-topological solitons, a self-interacting potential $U$ is needed, which we defined

\be
U\left( \bar{B}^{\mu} {B}_{\mu} \right) = m^2 \bar{B}^{\mu} {B}_{\mu} + \frac\lambda2  \left( \bar{B}^{\mu} {B}_{\mu} \right)^2 + \frac{h}{3} \left( \bar{B}^{\mu} {B}_{\mu} \right)^3 \, .
\label{potexp}
\ee
{ Here, $m$ is the mass of the boson, while $\lambda$ and $h$ are coupling constants.} The simplest spherically-symmetric ansatz that admits non-trivial radial profiles reads
\be
B= e^{i\omega t}\left( u(r) dt + i v(r) dr     \right)\,;\,\,\,\,\,\,\, A= A_t (r) dt\,.
\label{fieldansatz}
\ee

In this case, the equations of motion reduce to
\bea
\nn
 u''- \left( \omega+ q A_t     \right) \left( \frac2r v+ v'   \right)+ \frac2r u' - h u^5 + \left( g+2 h v^2   \right) u^3 -  \left( m^2 + g v^2 +h v^4 \right) u - q v A_t =0\,,\\
\left( -m^2 +\omega^2 + q A_t \left( 2\omega + q A_t  \right)    \right) v + \left(   \left( u^2 - v^2 \right)   \left( g - h \left(u^2 - v^2       \right)  \right)  \right)v -  \left( \omega+ q A_t     \right) u' =0\,,\\
\nn A_t'' + \frac2r A_t' +2 q v u' - 2 q  \left( \omega+ q A_t     \right) v^2 =0 \, .
\eea

Under these assumptions, it follows that near the origin the fields must behave as 
\bea
\nn
u(r)&\approx& u_{(0)} -\frac16 \left(\left(q A_{t(0)}+\omega\right)^2 + g u_{(0)}^2- h u_{(0)}^4 - m^2        \right) u_{(0)} r^2 + O(r^4)\,,\\
v(r)&\approx& -\frac13 \left(q A_{t(0)}+\omega\right)   u_{(0)} r + O(r^3)\,,\\ \nn
A_t(r)&\approx & A_{t(0)} - \frac1{90} q \left( g A_{t(0)}^2 -h A_{t(0)}^4 -m^2    \right) A_{t(0)}^2 \left( q A_{t(0)}+\omega \right) r^4+O(r^6)\, .
\eea

Then, in order to shoot to the desired asymptotic behavior at a fixed $\omega$, we can use the free parameters $u_{(0)}$, $ A_{t(0)}$ at the origin, naturally setting the constraint $\omega^2<m^2$, as the asymptotic behavior reads
\be
u(r)\rightarrow c_\infty e^{-\sqrt{m^2-\omega^2}}\,,\,\,\,\,\,\, v(r)\rightarrow \frac{c_\infty\omega}{-\sqrt{m^2-\omega^2}} e^{-\sqrt{m^2-\omega^2}}\,,\,\,\,\,\,\,\, A_t(r)\rightarrow \frac{A_\infty}{r}\,.
\ee

To ensure that the trivial solution is an absolute minimum of the energy $E=\int d^3x \sqrt{-g} \ \epsilon  =\int  d^3x \sqrt{-g}\ T_0^0$, the potential parameters must satisfy the relation
\be
h>\frac{\lambda^2}{4m^2}\,.
\ee

We are interested in solutions with a fixed value of the Noether charge $Q=\int d^3 x\sqrt{-g}\ \rho =\int d^3 x\sqrt{-g}\  j^0 $, where the non-topological solitons are the extrema of the $F=E-\omega Q$ functional.

\subsection*{Results}
\label{ballresults}

\begin{figure}[htp]
\begin{center}
\includegraphics[width=3.4in]{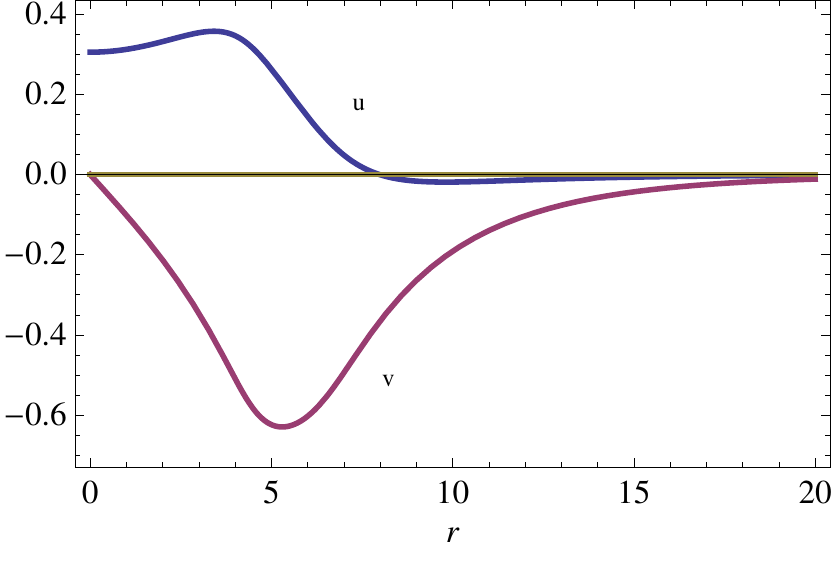}\hfill
\includegraphics[width=3.4in]{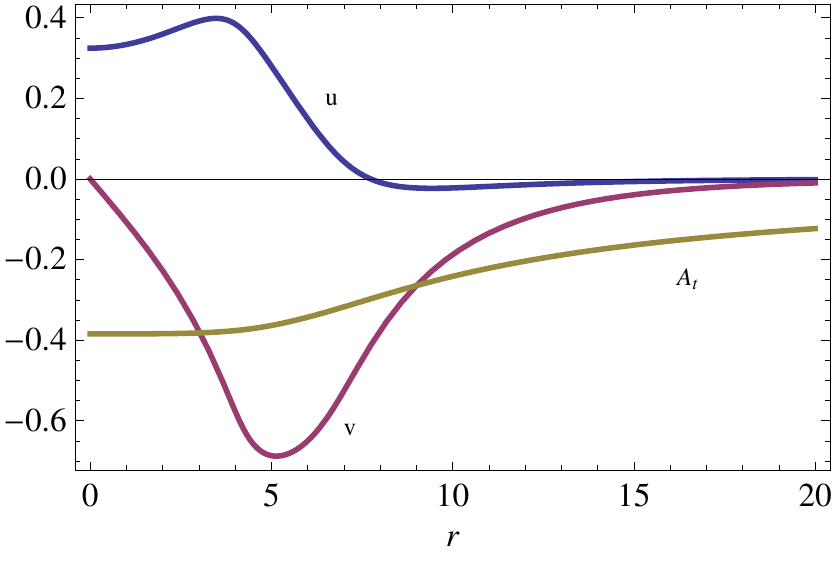}
\caption{\label{ballfiles} profiles for the uncharged (left panel) and charged (right panel, $q=0.02$) solutions for $\omega=0.980$.}
\end{center}
\end{figure}

Considering $m^2=-\lambda=h=1$ for the numerics, we found solutions with charges up to $q_{\rm max}=0.03$. On Figure~\ref{ballfiles} we show the field  profiles obtained for two such typical cases: $q=0$ and $q=0.02$.

\begin{figure}[htp]
\begin{center}
\includegraphics[width=3.4in]{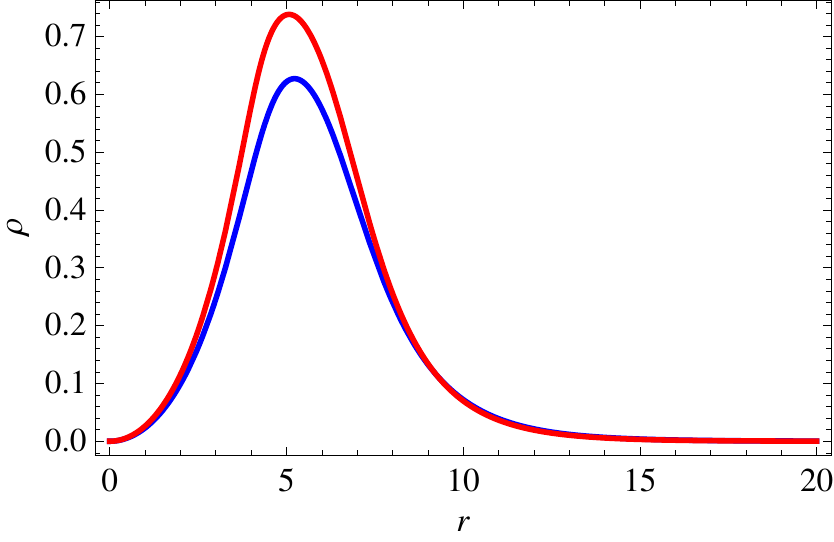}\hfill
\includegraphics[width=3.4in]{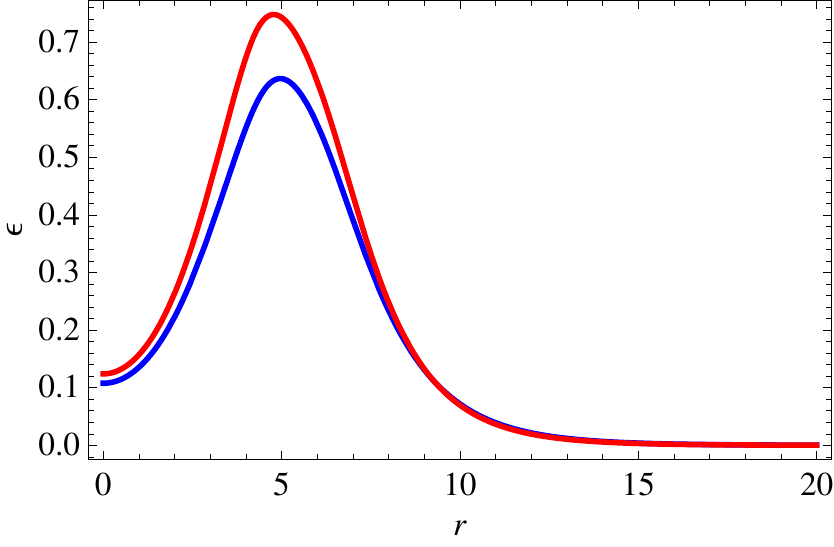}
\caption{\label{ballfilesre} Charge (left panel) and Energy (right panel) density profile for the charged ($q=0.02$, red) and uncharged (blue) solutions for $\omega=0.980$. }
\end{center}
\end{figure}

Writing explicity the charge and energy density, we obtain
\bea
\epsilon&=&u'^2+\omega^2 v^2-2\omega v u'+m^2 \left( u^2+v^2 \right)+\frac{\lambda}2 \left(  -3u^4+2u^2v^2+v^4        \right)+\frac{h}{3}\left(   u^2-v^2 \right)^2\left(5u^2+v^2\right)+\frac12 A_t'^2\,,\\
\rho&=&2 \left(\omega+qA_t\right) v^2- vu'\,.
\eea
On Figure~\ref{ballfilesre} we plot $\rho$ and $\epsilon$ profiles for the same cases described above.

\begin{figure}[htp]
\begin{center}
\includegraphics[width=3.4in]{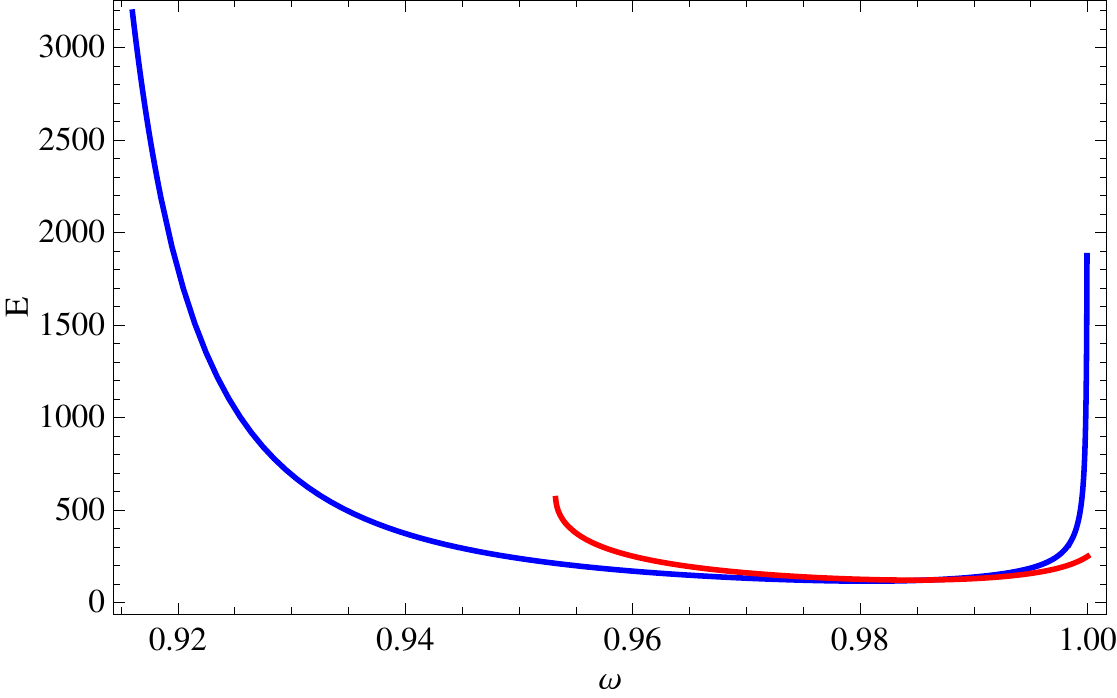}\hfill
\includegraphics[width=3.4in]{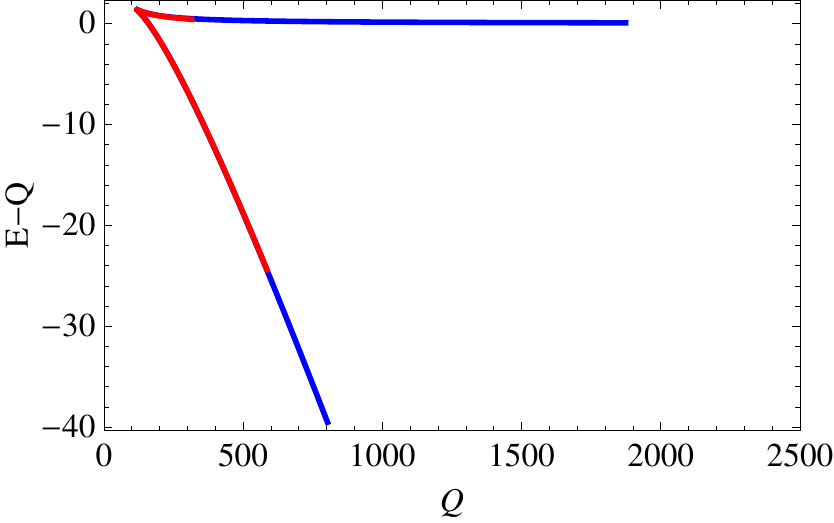}
\caption{\label{ballener} Left panel: energy as a function of the frequency for uncharged solutions with $q=0$ (blue) and charged solutions with $q=0.02$ (red). The total charge follows closely the energy curve. Right panel: difference between the energies corresponding to the soliton and the free-boson solution as a function of the total charge (for $m=1$). While both solitons are unstable to the decay of the free bosons in the thick-wall regime, they remain stable to this decay in the opposite thin-wall regime.}
\end{center}
\end{figure}

On the other hand, the dependences of $E$ and $Q$ are presented on Figure~\ref{ballener} as functions of the soliton frequency. On this Figure we present all the range for $\omega$ that we managed to reach by means of our numerical code. There are two particular limits at which the energy and the Noether charge tend to infinity. These limits correspond to the so-called thin-wall regime (when $\omega\rightarrow\omega_{\rm min}$) and the thick-wall regime (when $\omega\rightarrow1$). We found $\omega_{\rm min}\approx 0.915$ for $q=0$ and $\omega_{\rm min}\approx 0.953$ for $q=0.02$.


As the soliton solutions could still decay into free bosons, it follows that an important magnitude to be analyzed is the energy difference between these two solutions, i.e. $E-mQ$. The $Q$ dependence of this magnitude is shown in the right panel of Figure~\ref{ballener}. The curve obtained for $E-mQ$ consists of two branches, developing a spike at the junction point, where
the energy and the Noether charge of the soliton attain
their minimum values. This happens at $\omega\approx0.980$ for $q=0$ and $\omega\approx0.985$ for $q=0.02$. Roughly speaking, the plot shows that both charged and uncharged solitons are unstable to the decay in the free bosons in
the thick-wall regime meanwhile stable to this decay in the
thin-wall regime.

\section{Proca stars}
\label{stars}

In this section we focus on the study of spherically-symmetric static stars.
We analyze four different types of stars, depending on the choice of the values of the lagrangian parameters, that we name as follows:

\begin{itemize}

\item Proca stars (PS): this model was introduced in \cite{Brito:2015pxa} and corresponds to $q=0$, $m^2\neq0$, $\lambda=0$, $h=0$. In particular, we fix $m^2=1$.

\item Charged Proca stars (CPS): for which we gauge the $U(1)$ symmetry of the Proca stars by setting $q\neq0$. This is equivalent to the gauged scalar case studied in \cite{Jetzer:1989av} with respecto to the non-gauged works \cite{Kaup:1968zz,Ruffini:1969qy}. Just as in the scalar case, the maximum charge is $q^2=\frac12$. In particular, we take $q=0.5$, $m^2=1$, $\lambda=0$, $h=0$. 

\item Self-interacting Proca stars (IPS): these solutions are gravitating solutions  of Proca balls, introduced in \cite{Loginov:2015rya} and reviewed in Section~\ref{sballs} when setting $q=0$. We will also call them Proca Q-stars since they are the Proca version of Q-stars \cite{Lynn:1988rb}. In this case, for the numerics, we consider $m^2=-\lambda=h=1$.

\item Charged self-interacting Proca stars or charged Proca Q-stars (CIPS): the gauged version of the solutions described in the previous item. Their scalar version was studied in \cite{Prikas:2002ij}. Again, for the numerics, we assume $m^2=-\lambda=h=1$, $q=\frac12$.

\end{itemize}

For the gravitationally bounded solution, we start by assuming an ansatz for the metric
\be
ds^2=- \sigma^2(r) N(r)dt^2+ \frac1{N(r)}dr^2+r^2 d\Omega_2^2 \,.
\ee

As in the ``balls'' case studied in the previous Section, we consider the ansatz (\ref{fieldansatz}) for the matter fields and the expression (\ref{potexp}) for the potential.

In this case, the derived equations of motion read
\bea
\nn
\frac{d}{dr}\left( \frac{ \left(u'- \left(\omega + q A_t   \right)  r^2 \right)}{\sigma}\right)-\frac{r^2 u}{N^3\sigma^5}\left(-hu^4+\left(g+2hNv^2\right)Nu^2\sigma^2- \left( m^2+\left( g+h N v^2  \right)N v^2\right)N^2 \sigma^4   \right)&=&0\,,\\
\nn
 \left(\omega + q A_t   \right)u' + \left( m^2 +\left( g+h N v^2\right) N\sigma^2v^2 \right)N\sigma^2v- \left(\omega + q A_0   \right)^2v - \frac{1}{N\sigma^2}\left( -h u^4v+ \left(g+2 h N v^2 \right) \sigma^2      Nu^2 v \right)&=&0\,,\\
\nn
A_t''+\left(\frac2r - \frac{\sigma'}{\sigma} \right)A_t'+ 2 q v u' - 2 q  \left(\omega + q A_t   \right) v^2&=&0\,,\\
N'+\frac12 g r N^2 v^4+ \frac13 h r N^2 v^6+ \frac{rA_t'^2}{2\sigma^2}+\frac{ru'^2}{\sigma^2}+\frac{rv}{\sigma^2}\left( \left(  \left(\omega + q A_t   \right)^2 + m^2 N \sigma^2  \right) v-2  \left(\omega + q A_t   \right) u'   \right)&+&\\
\nn
+\frac{ru^2}{6N^3\sigma^6}\left(  10 h u^4 - 9 \left( g+2h N v^2 \right) N u^2\sigma^2     \right)+\frac{ru^2}{N\sigma^2} \left(m^2  +    \left(g+h N v^2     \right)Nv^2    \right) &=&0\,,
\\
\nn
\frac{\sigma'}{\sigma}+\frac{r}{2N\sigma^2}\left( \frac12 A_t'^2+\frac{\sigma^2N'}{r}   \right)+ \frac{N-1}{2rN}+\frac{r}{2N\sigma^2}\left( u' - \left(\omega + q A_t   \right) \right)^2+
\frac12 m^2 r \left(-v^2 - \frac{u^2}{N^2 \sigma^2}\right)&+&
\\
\nn
+\frac{gr}{4N^3\sigma^4}\left( u^4 + 2 N^2 u^2 v^2 \sigma^2 - 
 3 N^4 v^4 \sigma^4\right)-\frac{rh}{6N^4\sigma^6}\left(   u^2 - N^2 v^2 \sigma^2   \right)^2\left(u^2 + 5 N^2 v^2 \sigma^2\right)&=&0\,.
\eea

Near the origin, the solutions must behave as
\bea
\nn
u &\approx& u_c-\frac{u_c}{6 \sigma_c}\left( -h u_c^4+ \sigma_c \left( -m^2 \sigma_c^2 + \lambda u_c^2+   \left(\omega + q A_{tc}   \right)^2 \right)     \right)r^2 + O(r^4)\,,\\
\nn
v&\approx &-\frac{u_c}{3\sigma_c^2} \left(\omega + q A_{tc}   \right)^2 r +O(r^3)\,,
\\
\sigma&\approx & \sigma_c + \frac{1}{2\sigma_c^5}\left(m^2\sigma_c^4u_c^2-\sigma_c^2\lambda u_c^4+hu_c^6\right)r^2 + O(r^4) \,,\\
\nn
N&\approx&1-\frac1{18\sigma_c^6}\left( 6 m^2 \sigma_c^4 u_c^2 - 9\sigma_c^2 \lambda u_c^4 + 10 h u_c^6 \right) r^2 + O(r^4) \,,\\
\nn
A_t&\approx& A_{tc}+\frac{q u_c^2}{90\sigma^6}\left( m^2 \sigma_c^4-\lambda \sigma_c^2 u_c^2 + h u_c^4 \right)\left(\omega + q A_{tc}   \right) r^4 + O(r^6)\,.
\eea
Here we have three free parameters $u_c$, $\sigma_c$, $A_{tc}$ which we shall use to shoot into the proper behaviors at infinity: that is, regularity for the matter fields $u$, $v$ and the gauge field $A_t$; and asymptotic flatness $\sigma(r\rightarrow\infty)\rightarrow 1$.

\subsection*{Results}
\label{starsresults}

\begin{figure}[h]
\begin{center}
\includegraphics[width=3.4in]{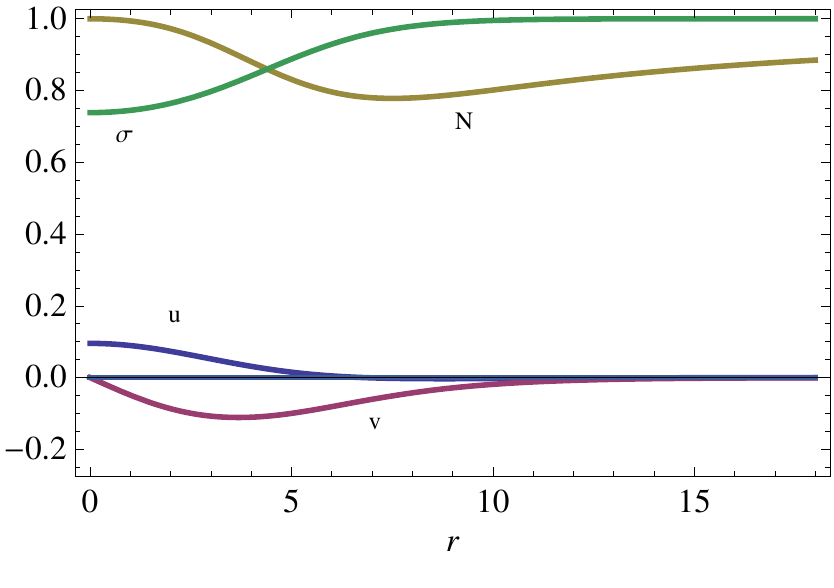}\hfill
\includegraphics[width=3.4in]{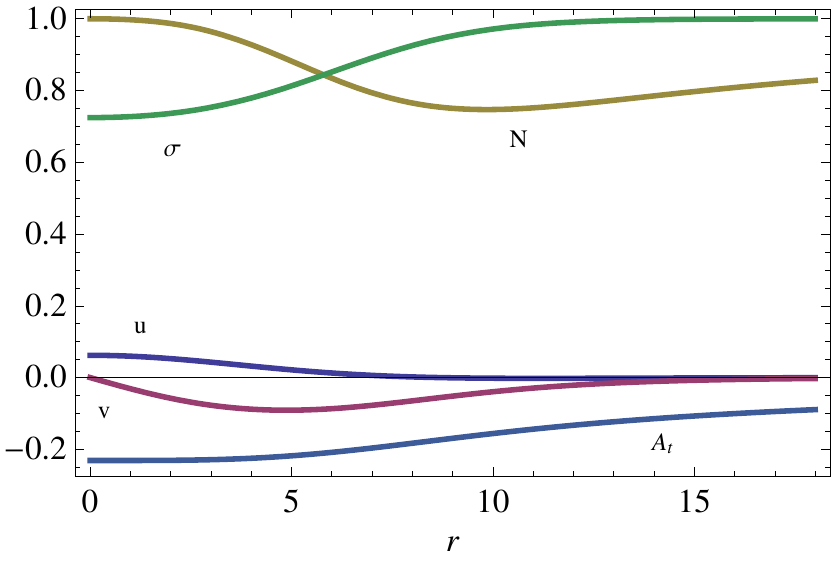}
\caption{\label{starfile}  Left panel: profiles of an interacting Proca star with $\omega=0.876$. Right panel: profiles of a charged interacting Proca star of $\omega=0.926$.}
\end{center}
\end{figure}

\begin{figure}[h]
\begin{center}
\includegraphics[width=3.4in]{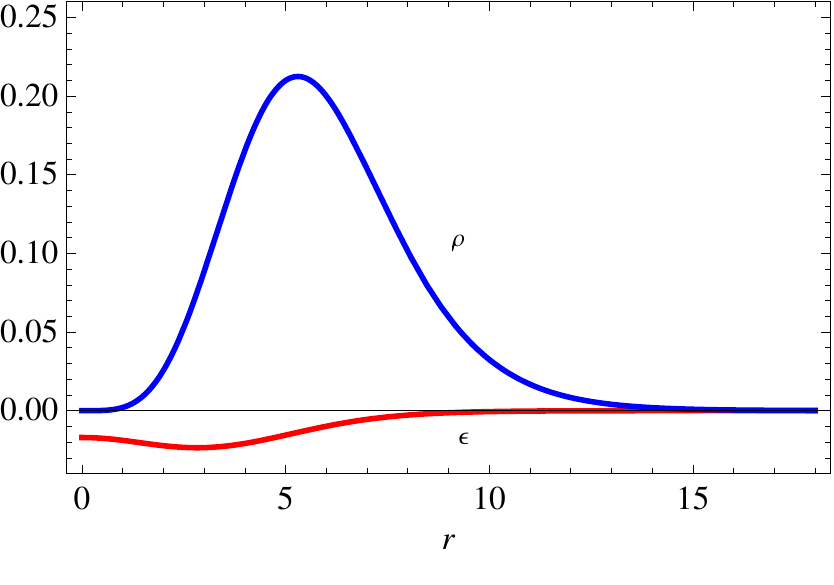}\hfill
\includegraphics[width=3.4in]{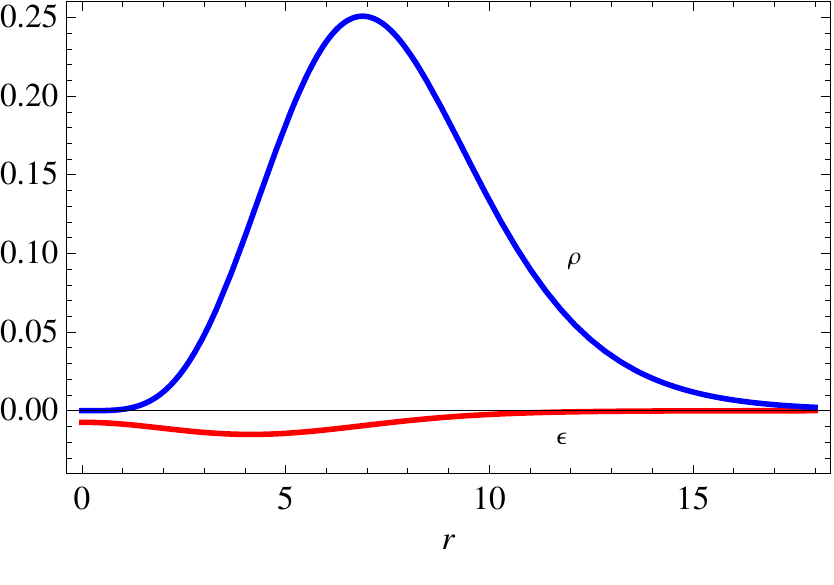}
\caption{\label{nederfile} Charge and energy-density profiles of the interacting Proca star of $\omega=0.876$ (left panel) and the charged interacting Proca star of $\omega=0.926$ (right panel).  }
\end{center}
\end{figure}

Considering the values of $m$, $q$, $\lambda$, and $h$ described above for the numerics, we found different sets of stars. On Figure~\ref{starfile} we show the field profiles obtained for two typical cases: an interacting Proca star with $\omega=0.876$ and a charged Proca star with $\omega=0.926$. Later, on Figure~\ref{nederfile} we plot $\rho$ and $\epsilon$ profiles for the same cases.

\begin{figure}[h]
\begin{center}
\includegraphics[width=3.4in]{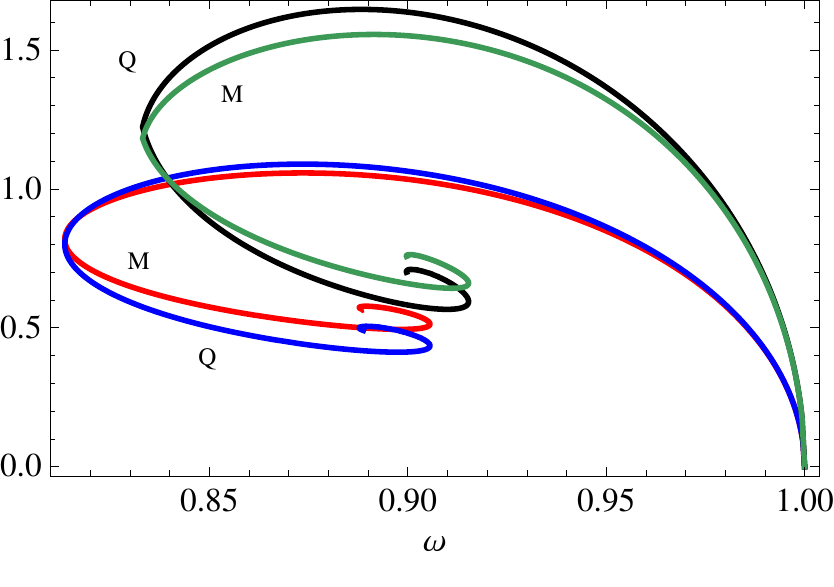}\hfill
\includegraphics[width=3.4in]{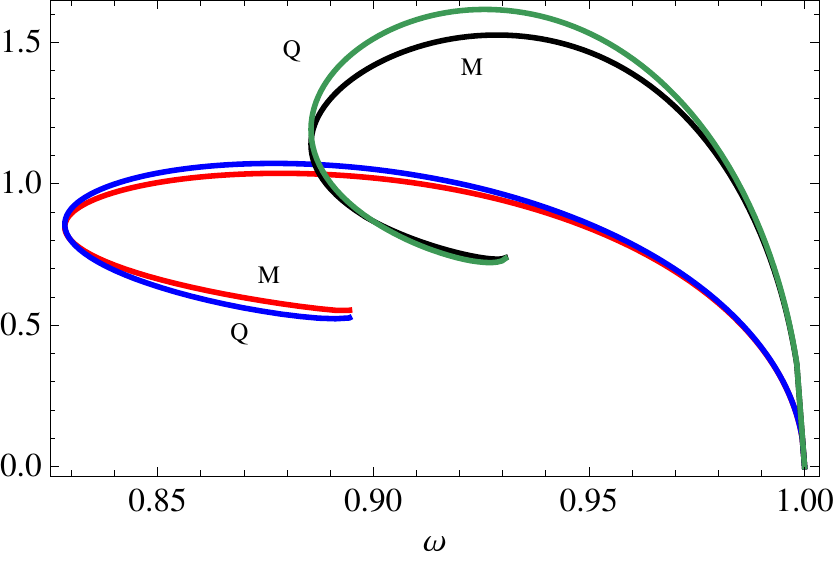}
\caption{\label{starfreq}  Left panel: ADM mass $M$ and Noether charge $Q$ for the Proca star (red and blue, respectively) and the charged Proca star (green and black, respectively) with respecto to the Proca field frequency $\omega$. Right panel: same magnitudes for the self-interacting case.  }
\end{center}
\end{figure}

On the left panel of Figure~\ref{starfreq} we plot the ADM masses $M$ and Noether charges $Q$ obtained for a whole set of uncharged ($q=0$) and charged ($q=0.5$) Proca stars considering field frequencies in the $\omega \sim 0.80-1.0$ range. For all cases, as $\omega \rightarrow 1$, both $M$ and $Q$ of the solutions vanish, while $Q/M \rightarrow 1$. In this limit, Proca stars become large and light, with very low mean densities, resulting trivial at $\omega = 1$. On the opposite, for smaller $\omega$, Proca stars get more compact. Both in the uncharged and charged cases, $M$ and $Q$ follow a spiral, towards different central configurations regarding on the choice of $q$. For $q=0$, this critical configuration is located around $\omega \approx 0.89$ \citep{Brito:2015pxa}, while for $q>0$ this critical value becomes larger. Moreover, while for $q=0$, the maximum of both $M$ and $Q$ occurs at $\omega_{\rm max} \approx 0.875$, with $M_{\rm max} < Q_{\rm max} \approx 1$, for $q>0$ all the critical values increase. Regarding the stability of these solutions, in the lower part of the spirals $M > Q$, and thus the binding energy $E = 1 - M/Q$ becomes negative, making all these regions unstable against perturbations \citep{Brito:2015pxa}. An analogue behaviour is found for the self-interacting cases, which follow similar trends in all cases, as can be seen on the right panel of Figure~\ref{starfreq}.

\begin{figure}[h]
\begin{center}
\includegraphics[width=3.4in]{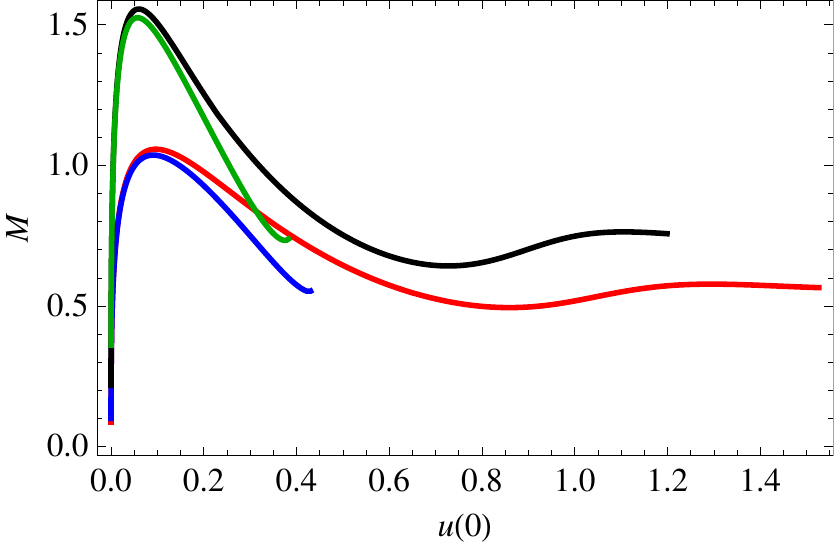}\hfill
\includegraphics[width=3.4in]{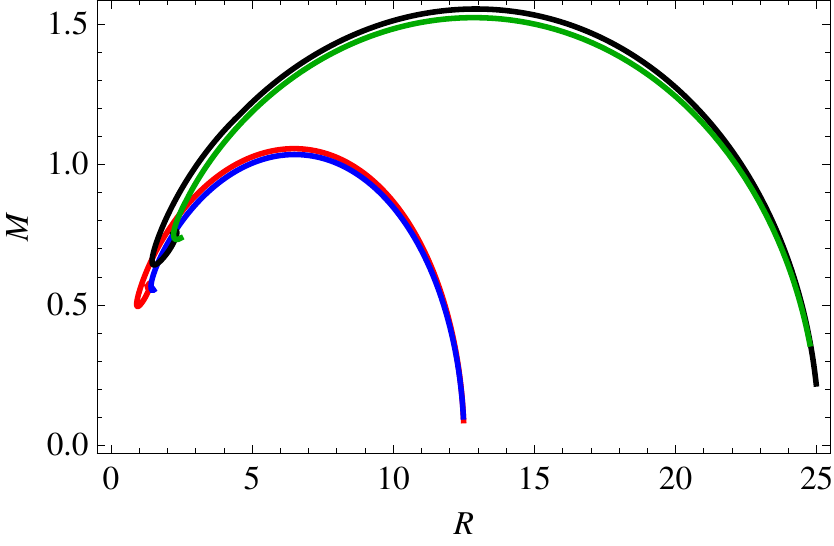}
\caption{\label{starmass}  ADM mass $M$ as a function of the central density $u(0)$ (left panel) and the radius $R$ (right panel) for the Proca star (red), the charged Proca star (black), the self-interacting Proca star (blue) and the charged self-interacting Proca star (green). }
\end{center}
\end{figure}


Finally, on the left panel of Figure~\ref{starmass} we show the total ADM mass $M$ as function of the central density $u(0)$ for the four cases considered in this Section: (un)charged Proca stars and (un)charged self-interacting Proca stars.
 On the right panel of Figure~\ref{starmass} we plot the mass-radius profiles of the same solutions.  
In order to do so, we define the radio of the Proca star as \cite{Jetzer:1989av}
\bea
R=\int d^3 x \sqrt{-g}\, r\, j^0
\eea
While uncharged and charged versions of both Proca and self-interacting stars show very similar trends in the mass-radius diagrams for the same parameters, they show a very different dependence on the central density, as the left panel of Figure~\ref{starmass} evidences. In both families, the gauged-stars versions are more massive and less compact than the uncharged ones.

\begin{table}[]
\centering
\caption{Maximum  masses and minimum radii for the different solutions expressed firstly in dimensionless units (for $m=1$) and then in physical units of solar masses $[M_\odot]$ and kilometers [km] for $m=10^{-10}$ eV.}
\label{table}
\begin{tabular}{l|c|c|c|c|c|c|c|c|}
\cline{2-9}
                                    & \multicolumn{2}{c|}{PS}       & \multicolumn{2}{c|}{IPS} & \multicolumn{2}{c|}{CPS} & \multicolumn{2}{c|}{CIPS} \\ \hline
\multicolumn{1}{|l|}{$m$}           & $M_{\rm max}$ & $R_{\rm min}$  & $M_{\rm max}$ & $R_{\rm min}$  & $M_{\rm max}$ & $R_{\rm min}$  & $M_{\rm max}$ & $R_{\rm min}$    \\ \hline
\multicolumn{1}{|l|}{$m=1$}       &      1.05         &         6.92      &     1.04        &    6.50        &   1.56        &      12.88        &     1.52      &      12.96        \\
\multicolumn{1}{|l|}{$10^{-10}$ eV} &     1.40     M$_\odot$      &        13.69 km       &      1.39  M$_\odot$      &     12.86 km       &    2.09  $M_\odot$      &     25.48  km       &     2.03  $M_\odot$      &    25.64 km          \\ \hline
\end{tabular}
\end{table}

We worked in natural units but also setting $m=1$. In order to recover physcal units  a physical value for the particle mass must be assumed. In Table \ref{table} we show maximum masses and minimum radii for a canonical case. Note that both the mass and the radius are meassured in units of $[M_{pl}^{2}/m]$ where $M_{pl}$ is the Planck mass. Then the compactness $\eta=\frac{2 M}{R}$ is independent of the particle mass (for each of the families considered). For our solutions we find $\eta_{\rm max}=$ 0.30 (PS), 0.32 (IPS), 0.24 (CPS), 0.23 (CIPS) which are of course less than the Schwarzschild black hole value $\eta=1$.

\section{Conclusions}
\label{conc}

We have studied a broad class of spherically symmetric regular solutions involving a complex massive vector. 

Firstly, we studied Proca-Q-balls, that is, non-gravitating solutions with a self-interaction potential which acts to hold the system together. We found both charged and uncharged solutions of this particular class of Q-balls below a critical charge depending on the choice of the free parameters available. For both charged and uncharged cases we found that solutions are only allowed in a limited range of frequencies $\omega_{min}<\omega<m$, with divergent energy in both minimum and maximum frequencies.
This extrema define the so called thin-wall ($\omega\rightarrow\omega_{min}$) and thick-wall ($\omega\rightarrow m$) regimes. Moreover, analyzing in detail the $E-m Q$ relation, we conclude that in the thick wall regime the solutions are unstable and will decay into free ``procons" or free Proca states.

Secondly, we also studied Proca-stars and Proca-Q-stars, that is, self-gravitating solutions. In this part, we extended the results of \cite{Brito:2015pxa} for a broader class of potentials showing that self-gravitating solutions are indeed robust.
In this sense, since it is reasonable to assume that dark matter in the Universe could be composed of different kinds of fundamental entities, condensates as vector Proca-stars studied here, as scalar Boson stars analyzed elsewhere, represent another viable dark component.

Looking forward, next steps for this study could come from extending our spherically symmetric to allow for rotation, in order to investigate spinning Proca-Q-balls, which could be done by considering a vector version of spinning non-topological solitons made of scalar fields \cite{Volkov:2002aj,Benci:2010zz,Brihaye:2009dx,Kleihaus:2005me}. In the same direction, it would also be interesting to find solutions where these fields act like a Kerr black hole hair or clouds following the steps of \cite{Hod:2012px,Hod:2014baa,Benone:2014ssa,Herdeiro:2014pka,Herdeiro:2014goa,Herdeiro:2016tmi}. On the other hand, clouds may also exist surrounding static black holes \cite{Sampaio:2014swa}. Extending the solutions found in this paper for our self-interacting model would then be a promising future approach to this field too.

\subsection*{Acknowledgements}

We are grateful to the referee for his/her insightful comments.
I. S. L. and F. G. acknowledge support from
CONICET, Argentina. We would like to thank Raulo Arias for carefully reading this manuscript. ISL would like to thank Juli for hospitality and cakes during several stages of this project. ISL thanks the IFLP, ICTP and Galileo Galilei Institute for Theoretical Physics for hospitality. FG is grateful to Luz and Paz for cordial reception.

\end{document}